\documentclass[12pt]{article}
\usepackage{pic02}
\usepackage{hyperref}
\usepackage{url}
\usepackage{graphicx}

\begin{document}

\title{\bf TEST OF GENERAL RELATIVITY:
1995-2002 MEASUREMENT OF FRAME-DRAGGING}
\author{Ignazio Ciufolini        \\
{\em Dip. Ingegneria dell'Innovazione, Universit\`a di Lecce, Via
Monteroni, 73100 Lecce, Italy}} \maketitle

%
%
%
%
%
%
\vspace{4.5cm}
%

\baselineskip=14.5pt
\begin{abstract}
After an introduction on phenomena due to spin and mass-energy
currents on clocks and photons, we review the 1995-2001
measurements of gravitomagnetic field of Earth and Lense-Thirring
effect obtained by analyzing the orbits of the two laser-ranged
satellites LAGEOS and LAGEOS II; this method has provided a
direct measurement of Earth's gravitomagnetism with accuracy of
the order of 20 $\%$. A future accurate measurement of the
Lense-Thirring effect, at the level of 1$\%$ accuracy, may
include the LARES experiment that will also provide other basic
tests of general relativity and gravitation. Finally, we report
the latest measurement of the Lense-Thirring effect, obtained in
2002 with the LAGEOS satellites over nearly 8 years of data. This
2002 result fully confirms and improves our previous measurements
of the Earth frame-dragging: the Lense-Thirring effect exists and
its experimental value is within $\sim$ 20$\%$ of what is
predicted by Einstein's theory of general relativity.
\end{abstract}
\newpage

\baselineskip=17pt

\section{Gravitomagnetic phenomena on test gyroscopes,
test particles, clocks and photons}

In this paper we review some phenomena arising in the vicinities
of a rotating body and some proposals and recent measurements of
frame-dragging and Lense-Thirring effect obtained by laser-ranged
satellites. For a general review of tests and measurements of
general relativistic and gravitational effects we refer to
\cite{ciuw,wil1}.

Einstein's general theory of relativity \cite{ciuw} predicts
the occurrence of peculiar phenomena on test gyroscopes, test
particles, clocks and photons in the vicinity of a mass-energy
current and thus in the vicinity of a spinning body due to its
rotation.

In section (2) we describe how the orbit of a test-particle is
influenced by the spin of a central body and the direct
measurement of the orbital perturbations of laser ranged
satellites due to the Earth spin, i.e. the Lense-Thirring effect
\cite{len}. In 1995-2002 the Lense-Thirring effect was measured
with about 20$\%$ accuracy using the LAGEOS and LAGEOS II
satellites \cite{ci1,ci2,ci3,ci4}, see section (2).

Small test-gyroscopes, that determine the axes of a local, freely
falling, inertial frame, where ``locally'' the gravitational
field is ``unobservable'', rotate with respect to ``distant
stars'' due to the spin of a body. This effect is described in
\cite{ciuw} and should be measured with about 1$\%$ accuracy by
the Gravity Probe-B experiment  \cite{eve}.

However, not only test particles and gyroscopes are affected by
the spin of the central object but also photons and clocks. In
this section we review some gravitomagnetic phenomena on clocks
and photons. A photon co-rotating around a spinning body takes
less time to return to a ``fixed point'' (with respect to distant
stars) than a photon rotating in the opposite direction
\cite{ciuw,ciur1}. Since light rays are used to synchronize
clocks, the different travel-time of co-rotating and
counter-rotating photons implies the impossibility of
synchronization of clocks all around a closed path around a
spinning body; the behavior of light rays and the behavior of
clocks around a spinning body are intimately connected. In
several papers the ``frame-dragging clock effect'' around a
spinning body has been estimated and some space experiments have
been proposed to test it \cite{mas,tar,ciur1}. Thus, when a
clock, co-rotating very slowly (using rockets) around a spinning
body and at a constant distance from it, returns to its starting
point, it finds itself advanced relative to a clock kept there at
``rest'' (with respect to ``distant stars'', see above).
Similarly a clock, counter-rotating arbitrarily slowly and at a
constant distance around the spinning body, finds itself retarded
relative to the clock at rest at its starting point
\cite{ciuw,ciur1}. For example, when a clock that co-rotates very
slowly around the spinning Earth, at $r \, \sim$ 6000 km altitude,
returns to its starting point, it finds itself advanced relative
to a clock kept there at ``rest'' (with respect to ``distant
stars'') by $~ \Delta \tau ~ \sim {{\frac{4 \, \pi \,
J_\oplus}{r}}} \sim 5 \times 10^{-17}$ $sec$, where $J_\oplus
\cong \, 145 \, cm^2$ is the Earth angular momentum. Similarly, a
clock, that counter-rotates very slowly around the spinning
Earth, finds itself retarded relative to a clock kept there at
``rest'' by the same amount. Then, the difference between the
time read by the two clocks when they meet again after a whole
revolution is about $\sim 10^{-16}$ \cite{ciur1,tar}.

However, Einstein's gravitational theory predicts peculiar
phenomena also inside a rotating shell \cite{ciuw}.
In reference \cite{ciur1} we derive the time-delay in
travel-time of photons due to the spin of a body both outside a
rotating body and inside a rotating shell. We then show that this
time-delay by the spin of an astrophysical object might be
detected in different images of the same source by gravitational
lensing.

Since here we are only interested to analyze the time delay due
to spin, we chose a simple configuration where source, lens and
observer are aligned and we use quasi-Cartesian coordinates
\cite{ciur1}. We then get, for a photon with impact parameter $b$
traveling on the equatorial plane of the source: $\Delta T =
4\,M\ln \left( {2\frac{\bar z} {b}} \right) + \frac{4\,J} {b}$.
In this expression $\bar z \gg b$ is the distance of source and
observer from the lens, the first term is the standard Shapiro
time delay and the second term is the gravitomagnetic time-delay
due to the spin of the deflecting body.

Let us now give the order of magnitude of the time delay due to
the spin of some astrophysical sources. For the sun, by
considering two light rays on the equatorial plane of the Sun,
grazing the Sun on opposite sides, the relative gravitomagnetic
time delay is $\Delta T_{rel}^J=3.35\cdot 10^{-12}\,$ sec. For
the lensing galaxy of the Einstein cross, by assuming a simple
model for rotation and shape of the central object (see
\cite{ciur1} and references therein), we then get: $\Delta
T_{12}^J \simeq8\,hr$. As a third example we consider the
relative time delay of photons due to the spin of a typical
cluster of galaxies. Depending on the geometry of the system and
on the path followed by the photons, we then find relative time
delays ranging from a few minutes to several days. Then, at least
in principle, one could detect the time delay due to the spin of
a lensing galaxy by removing the larger quadrupole-moment time
delay by a method described in \cite{ciur1}; of course, as in the
case of the Sun, one should be able to accurately enough model
and remove all the other delays, due to other physical effects,
from the observed time delays between the images.

Let us now analyze the time delay in the travel time of photons
propagating inside a shell of mass $M$ rotating with angular
velocity $\omega$. Inside a spinning shell it is in general not
possible to synchronize clocks all around a closed path. Indeed,
if we consider a clock co-rotating very slowly on the equatorial
plane along a circular path with radius $r$, when back to its
starting point it is advanced with respect to a clock kept there
at rest (in respect to distant stars). The difference between the
time read by the co-rotating clock and the clock at rest is equal
to: $\delta T =\frac{{8M}} {3R}\pi \omega r^2$.
 For a shell with
finite thickness we just integrate this expression from the
smaller radius to the larger one.

If we now consider a photon traveling with an impact parameter $r$
on the equatorial plane of a galaxy; the time delay due to the
rotation of the external mass for every infinitesimal shell with
mass $dm=4\pi \rho R'^2 dR'$ and radius $R'\geq \left| r
\right|$, is \cite{ciur1}: $\Delta t_{dm}  = \frac{{8\,dm}}
{3}\omega r\frac{{\sqrt {R'^2 - r^2 } }} {{R'}}$. This is the
time delay due to the spin of the external thin shell. By
integrating this expression from ${\left| r \right|}$ to the
external shell radius $R$, we have:

\begin{equation}\label{1}
\Delta T =\frac{{32\pi }} {3}\omega r\int_{\left| r \right|}^R
{\rho R'} \sqrt {R'^2  - r^2 } dR'
\end{equation}

This is the time delay due to the spin of the whole rotating mass
of the external shell. From this formula we can easily calculate
the relative time delay between two photons traveling on the
equatorial plane of a rotating shell, with impact parameters
$r_1$ and $r_2$.

Let us give the time delay corresponding to some astrophysical
configurations. In the case of the "Einstein cross", after some
calculations, based on a standard model for the lensing galaxy
(see ref. \cite{ciur1} and references therein), the order of
magnitude of the relative time delay of two photons traveling at
a distance of $r_1\simeq650\,pc$ and $r_2\simeq - 650\,pc$ from
the center, using (\ref{1}) in the case $r _1\simeq - r_2$, is:
$\Delta T\simeq 20\,$ min. If the lensing galaxy is inside a
rotating cluster, or super-cluster, to get an order of magnitude
of the time delay due to the spin of the mass rotating around the
deflecting galaxy, we use typical super-cluster parameters (see
\cite{ciur1} and references therein). If the galaxy is in the
center of the cluster and light rays have impact parameters $r_1
\simeq 15\, kpc$ and $r_2 \simeq - 15\, kpc$ (of the order of the
Milky Way radius), the time delay, applying formula (\ref{1}) in
the case $r_1 \simeq - r_2$ and $\rho= constant$, is: $\Delta
t\simeq 1 \, day$.

Finally, if the lensing galaxy is not in the center of the
cluster but at a distance $r = a R$ from the center, with $0 \leq
a \leq 1$ and $R$ radius of the cluster, by integrating (\ref{1})
between $r = a R$ and $R$, when $r_1 \simeq r_2$ we have:

\begin{equation}\label{2}
\Delta T = \frac{32\pi}{9} \omega ({r_1  - r_2}) \rho {(1 -
a^2)^{1/2}} (1 - 4 a^2) R^3
\end{equation}

Thus, if the lensing galaxy is at a distance of 10 Mpc from the
center of the cluster, the relative time delay due to the spin of
the external rotating mass between two photons with
$(r_1-r_2)\simeq 30\, kpc$, is: $\Delta T\simeq 0.9 \, day$.
Since the present measurement uncertainty in the lensing time
delay is of the order of 0.5 day \cite{big}, the spin time delay
might already be observable.

In conclusion, in ref. \cite{ciur1} we have derived and studied
the "{\it spin-time-delay}" in the travel time of photons
propagating near a rotating body, or inside a rotating shell due
to the angular momentum. We found that there may be an
appreciable time delay due to the spin of the body, or shell,
thus {\it spin-time-delay} must be taken into account in the
modeling of relative time delays of the images of a source
observed at a far point by gravitational lensing. This effect is
due to the propagation of the photons in opposite directions with
respect to the direction of the spin of the body, or shell. If
other time-delays can be accurately enough modeled and removed
from the observations \cite{ciur1}, one could directly measure the
spin-time-delay due to the gravitomagnetic field of the lensing
body. We have analyzed the relative time-delay in the
gravitational lensing images caused by a typical rotating galaxy,
or cluster of galaxies. We have then analyzed the relative
spin-time-delay when the path of photons is inside a galaxy, a
cluster, or super-cluster of galaxies rotating around the
deflecting body; this effect should be large enough to be
detected at Earth. The measurement of the spin-time-delay, due to
the angular momentum of the external massive rotating shell,
might be a further observable for the determination of the total
mass-energy of the external body, i.e. of the dark matter of
galaxies, clusters and super-clusters of galaxies. Indeed, by
measuring the spin-time-delay one can determine the total angular
momentum of the rotating body and thus, by estimating the
contribution of the visible part, one can determine its
dark-matter contribution. The estimates presented in this paper
are preliminary because we need to apply the spin-time-delay to
some particular, known, gravitational-lensing images.
Furthermore, we need to estimate the size and the possibility of
modeling other sources of time-delay. Nevertheless, we conclude
that, depending on the geometry of the astrophysical system
considered, the relative spin-time-delay can be a quite large
effect.

\section{Measurement of gravitomagnetism with laser ranged satellites}

In ref. \cite{ciup}  (see also ref. \cite{ci5}) we describe the
LARES experiment to measure the Lense-Thirring effect with
relative accuracy of about 1$\%$. This laser ranged satellite, by
detecting its perigee rate, would also test the foundations and
other basic phenomena of general relativity and gravitational
interaction. Indeed, LARES would improve the bounds on
hypothetical long-range gravitational forces and the bounds on
deviations from the inverse square law for very weak-field
gravity; LARES would improve, by about two orders of magnitude,
the accuracy in testing the equivalence principle and would
provide an improved measurement in the field of Earth of the PPN
(Parametrized-Post-Newtonian) parameters $\alpha_1, \, \beta \,$
and $\gamma$ \cite{ciup}.

In this
section we describe the 1995-2002 measurements of the
Lense-Thirring effect obtained using LAGEOS and LAGEOS II.

The measurement of distances has always been a fundamental issue
in astronomy, engineering, and science in general.  So far, laser
ranging has been the most accurate technique for measuring the
distances to the moon and artificial satellites. Short-duration
laser pulses are emitted from lasers on Earth, aimed at the
target through a telescope, and then reflected back by optical
cube-corner retroreflectors on the moon or an artificial satellite
\cite{de}, such as LAGEOS.  By measuring the total round-trip
travel time, one can determine the distance to a retroreflector
on the moon with an accuracy of about 2 cm and to the LAGEOS
satellites with a few millimeters accuracy.

Our detection and measurement of the Lense-Thirring effect was
obtained by using the satellite laser ranging data of the
satellites LAGEOS (LAser GEOdynamics Satellite, NASA) and LAGEOS
II (NASA and ASI, the Italian Space Agency) and the Earth
gravitational field models, JGM-3 and EGM-96. The LAGEOS
satellites are heavy brass and aluminum satellites, of about 406
kg weight, completely passive and covered with retroreflectors,
orbiting at an altitude of about 6,000 km above the surface of
Earth.  LAGEOS, launched in 1976 by NASA, and LAGEOS II, launched
by NASA and ASI in 1992, have an essentially identical structure
but they have different orbits.  The semimajor axis of LAGEOS is
$a \, \cong \, 12,270$ km, the period $P \, \cong \, 3.758$ ~hr,
the eccentricity $e \, \cong \, 0.004$, and the inclination $I \,
\cong \, 109.9^{\circ}$.  The semimajor axis of LAGEOS II is
$a_{II} \, \cong \, 12,163$ km, the eccentricity $e_{II} \, \cong
\, 0.014$, and the inclination $I_{II} \, \cong \, 52.65^{\circ}$.

We analyzed the laser-ranging data adopting the IERS
(International Earth Rotation Service) conventions \cite{mc} in
our modeling, however, in the 1998 analysis, we used the static
and tidal EGM-96 model \cite{le}.  Error analysis of the LAGEOS
orbits indicated that the EGM-96 errors can only contribute
periodic root-sum-square errors of 2 to 4 mm radially, and in all
three directions they do not exceed 10 to 17 mm.  The initial
positions and velocities of the LAGEOS satellites were adjusted
for each 15-day batch of data, along with variations in their
reflectivities.  Solar radiation pressure, Earth albedo, and
anisotropic thermal effects were also modeled \cite{ru1}. In
modeling the thermal effects, the orientation of the satellite
spin axis was obtained from ref. \cite{fa}.  Lunar, solar, and
planetary perturbations were also included in the equations of
motion, formulated according to Einstein's general theory of
relativity with the exception of the Lense-Thirring effect, which
was purposely set to zero.  All of the tracking station
coordinates were adjusted (accounting for tectonic motions)
except for those defining the TRF terrestrial reference frame.
Polar motion was also adjusted, and Earth's rotation was modeled
from the very long baseline interferometry-based series SPACE96
\cite{gr}.  We analyzed the orbits of the LAGEOS satellites using
the orbital analysis and data reduction software GEODYN II
\cite{pa}.

The node and perigee of LAGEOS and LAGEOS II are dragged by the
Earth's angular momentum. From the Lense-Thirring formula
\cite{ci1,ci2,ci3}, we get $ \dot { \Omega}^{Lense-Thirring}_{I}$
$\cong 31$ mas/yr and $ \dot { \Omega\/}^{Lense-Thirring}_{II} \,
\cong 31.5$, where mas is a millisecond of arc. The argument of
pericenter (perigee in our analysis), $\omega$, also has a
Lense-Thirring drag \cite{ciuw}, thus, we get for LAGEOS: $ \dot {
\omega \/}^{Lense-Thirring}_{I} ~ $ $ \cong ~ 32$ mas/yr, and for
LAGEOS II: $ \dot { \omega \/}^{Lense-Thirring}_{II} ~ $ $ \cong
~ - 57$ mas/yr \cite{ci1,ci2,ci3}. The nodal precessions of LAGEOS
and LAGEOS II can be determined with an accuracy of the order of 1
mas/year. Over our total observational period of about 4 years,
we obtained a root mean square (RMS) of the node residuals of
about 4 mas for LAGEOS and about 7 mas for LAGEOS II \cite{ci3}.
For the perigee, the observable quantity is the product $e \cdot
a \cdot {\dot \omega}$, where $e$ is the orbital eccentricity of
the satellite. Thus, the perigee precession $\dot \omega$ for
LAGEOS is difficult to measure because its orbital eccentricity
$e$ is $ \sim 4 \times 10^{-3}$. The orbit of LAGEOS II is more
eccentric, with $e \sim $ 0.014, and the Lense-Thirring drag of
the perigee of LAGEOS II is almost twice as large in magnitude as
that of LAGEOS. Over about 4 years, we obtained a root mean
square of the residuals of the LAGEOS II perigee of about 25 mas
\cite{ci3}, whereas the total Lense-Thirring effect on the perigee
is, over 4 years, $\cong \, -228$ mas.

To precisely quantify and measure the gravitomagnetic effects we
have introduced the parameter $\mu$ that is by definition 1 in
general relativity \cite{ciuw} and zero in Newtonian theory.

The main error in this measurement is due to the uncertainties in
the Earth's even zonal harmonics and their time variations.  The
unmodeled orbital effects due to the harmonics of lower order are
comparable to, or larger than, the Lense-Thirring effect.
However, by analyzing both the JGM-3 and the EGM-96 models with
their uncertainties in the even zonal harmonic coefficients and
by calculating the secular effects of these uncertainties on the
orbital elements of LAGEOS and LAGEOS II, we find \cite{ci1} that
the main sources of error in the determination of the
Lense-Thirring effect are concentrated in the first two even
zonal harmonics, $J_{2}$ and $J_{4}$.  We can, however, use the
three observable quantities $\dot \Omega_{I}$, $\dot \Omega_{II}$
and $\dot \omega_{II}$ to determine $\mu$ \cite{ci1}, thereby
avoiding the two largest sources of error, those arising from the
uncertainties in $J_{2}$ and $J_{4}$. We do this by solving the
system of the three equations for the three residuals $\delta \dot
\Omega_{I}$, $\delta \dot \Omega_{II}$ and $\delta \dot
\omega_{II}$ in the three unknowns $\mu$, $J_{2}$ and $J_{4}$,
obtaining:

\begin{displaymath}
{\delta \, \dot \Omega^{Exp}_{LAGEOS I} + c_{1} \, \delta \,
\dot \Omega^{Exp}_{LAGEOS II} + c_{2} \, \delta \, \dot
\omega^{Exp}_{LAGEOS II} \,} =
\end{displaymath}

\begin{equation}\label{3}
= {\, \mu \, ( \, 31 \, + \, 31.5 \, c_{1} \, - \, 57 \, c_{2})
\, mas/yr} \, + \, {other \,\, errors} \cong \mu \, (60.2 \,
mas/yr),
\end{equation}

where $c_1=0.295$ and $c_2= - 0.35$. Equation (\ref{3}) for $\mu$
does not depend on $J_{2}$ and $J_{4}$ nor on their
uncertainties; thus, the value of $\mu$ that we obtain is
unaffected by the largest errors, which are due to $\delta J_2$
and $\delta J_4$, and is sensitive only to the smaller errors due
to $\delta J_{2n}$, with $2n \geq 6$.

Similarly, regarding tidal, secular, and seasonal changes in the
geopotential coefficients, the main effects on the nodes and
perigee of LAGEOS and LAGEOS II, caused by tidal and other time
variations in Earth's gravitational field \cite{ci1,ci2,ci5}, are
due to changes in $J_{2}$ and $J_{4}$. However, the tidal errors
in $J_{2}$ and $J_{4}$ and the errors resulting from other
unmodeled, medium and long period, time variations in $J_{2}$ and
$J_{4}$, including their secular and seasonal variations, are
eliminated by our combination of residuals of nodes and perigee.
In particular, most of the errors resulting from the 18.6- and
9.3-year tides, associated with the lunar node, are eliminated in
our measurement. An extensive discussion of the various error
sources that can affect our result is given in \cite{ci1,ci2,ci5},
only a brief discussion of the error sources is given below.

Let us now report the main results of our measurements. In Fig.
1, we display the improved analysis \cite{ci3} (obtained with the
linear combination of the residuals of the nodes of LAGEOS and
LAGEOS II and perigee of LAGEOS II according to Eq. (\ref{3}))
using the static and tidal Earth gravitational model EGM-96, we
also refined the non-gravitational perturbations model, the total
period of observations was of 4 years, longer of about 1 year
with respect to the previous observational period corresponding
to ref. \cite{ci2}. We only removed three small periodic residual
signals and the small observed inclination residuals. The removal
of the periodic terms was achieved by a least squares fit of the
residuals using a secular trend and three periodic signals with
1044-, 820-, and 569-day periods, corresponding to, respectively,
the nodal period of LAGEOS, and the perigee and nodal periods of
LAGEOS II. The 820-day period is the period of the main odd zonal
harmonics perturbations of the LAGEOS II perigee; the 1044- and
569-day periods are the periods of the main tidal orbital
perturbations, with $l$=2 and $m$=1, which were not eliminated
using Eq. (\ref{3}). Some combinations of these frequencies
correspond to the main non-gravitational perturbations of the
LAGEOS II perigee. We notice that this analysis, using EGM-96 and
its accurate tidal model, is substantially independent of the
removed signals, whereas the previous analysis \cite{ci2}, was in
part sensitive on the periodic terms included in the fit.  In
other words, our value (Fig. 1) of the secular trend does not
significantly change by fitting additional periodic
perturbations, and indeed, even the fit of the residuals with a
secular trend only, with no periodic terms, just changes the slope
by about 10 $\%$. Nevertheless, in this case, the root mean square
of the post-fit residuals increases by about four times with
respect to Fig. 1.

\begin{figure}[htb]
\includegraphics[width=10cm]{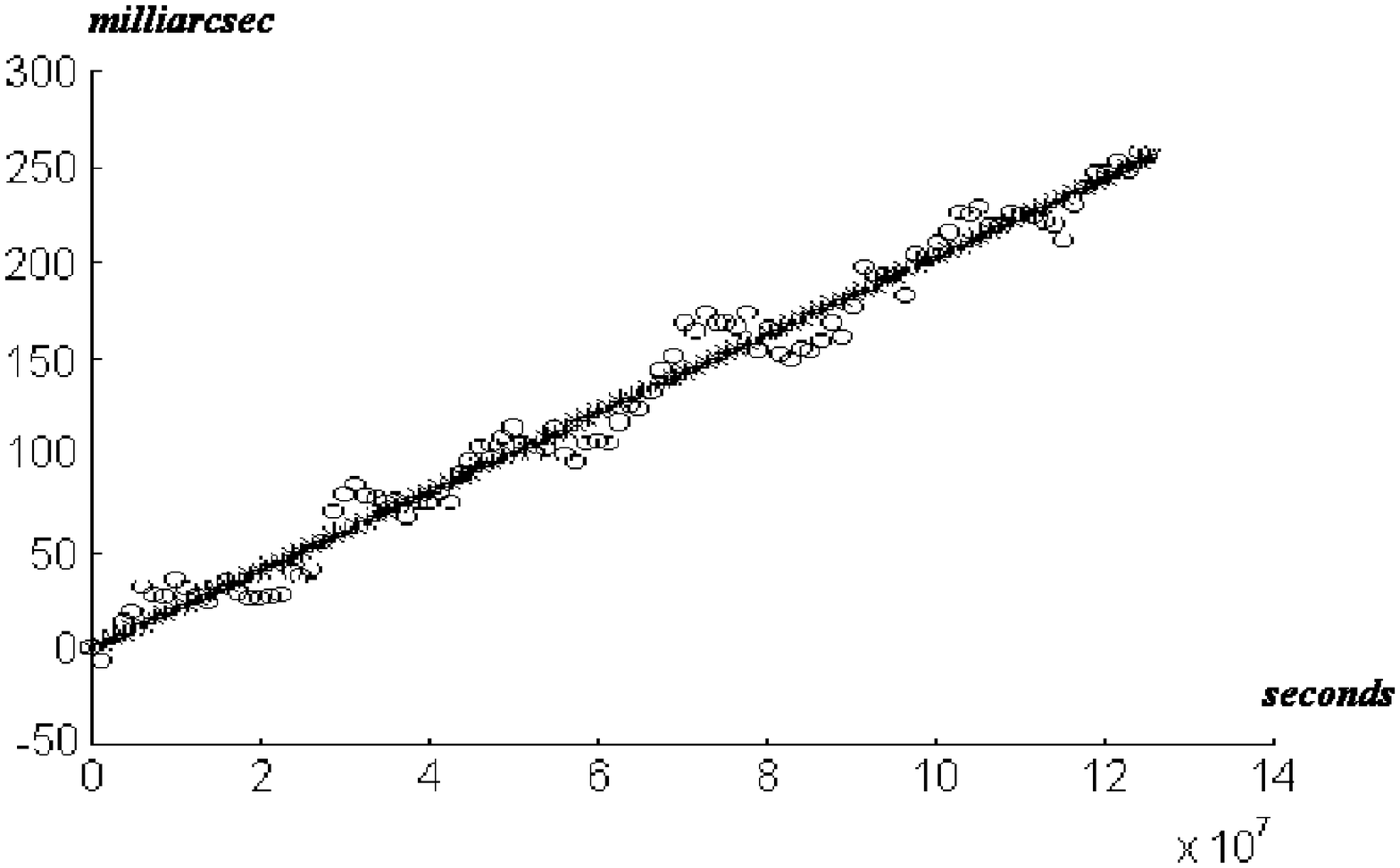}
 \caption{\it 1997-measurement of the Lense-Thirring effect.
Combination of the residuals of the nodes of LAGEOS and LAGEOS II
and perigee of LAGEOS II according to Eq. (\ref{3}), using the
Earth gravitational model EGM-96, over a 4-year period
\cite{ci3}.}
\end{figure}

Our best-fit straight lines of Fig. 1, through the combined
residuals of nodes and perigee, has a slope ${\mu}^{Measured}
\cong 1.1 \pm 0.03,$ where 0.03 is the standard deviation of the
fit. This combined, measured, gravitomagnetic perturbation of the
satellites' orbits corresponds, in a 4-year period, to about 16 m
at the LAGEOS altitude, that is, about 265 mas.

The root mean square of the post-fit combined residuals
corresponding to Fig. 1 is about 9 mas.  Our total systematic
error is estimated to be of the order of 30$\%$-40$\%$ of
$\mu_{GR}$ corresponding to the previous analyses of ref. \cite{ci2}, and of the order of 20$\%$-25$\%$ of $\mu_{GR}$
corresponding to Figure 1 \cite{ci3}.

Using the JGM-3 covariance matrix, we found the errors due to the
uncertainties in the even zonal harmonics $J_{2n}$, with $2n \geq
6$, to be:  $\delta \, {\mu}^{even \, zonals:  J_{2n} \geq J_6}
\,$ $ {\buildrel < \over \sim} 17 \, \%$ of $\mu_{GR}$, and using
the EGM-96 covariance matrix:  $\delta \, {\mu}^{even \, zonals:
J_{2n} \geq J_6} \, {\buildrel < \over \sim} 13 \, \%$ of
$\mu_{GR}$. The errors in the modeling of the perigee rate of
LAGEOS II due to the uncertainties in the odd zonal harmonics
$J_{2n+1}$ are, with EGM-96: $\delta \, {\mu}^{odd \, zonals} \,
{\buildrel < \over \sim} 2 \, \%$ of $\mu_{GR}$. Using the EGM-96
tidal model, we estimated the effect of tidal perturbations and
other variations of Earth gravitational field to be $\delta \,
{\mu}^{tides \, + \, other variations} \, {\buildrel < \over
\sim} 4 \, \%$ of $\mu_{GR}$. On the basis of analyses
\cite{ci5,luc} of the non-gravitational perturbations --- in
particular, those on the perigee of LAGEOS II --- we found
$\delta \, {\mu}^{non-gravitational} \, {\buildrel < \over \sim}
13\%$-20$\%$ of $\mu_{GR}$, including uncertainties in the
modeling of the satellites' reflectivities, and the error due to
uncertainties in the orbital inclinations of LAGEOS and LAGEOS II
was estimated to be $\delta \, {\mu}^{inclination} \, {\buildrel
< \over \sim} 5 \, \%$ of $\mu_{GR}$.

 Taking into account all of these error sources, we arrived at a
total root-sum-square error ${\buildrel < \over \sim} 20\%$-25$\%$
of $\mu_{GR}$. Therefore, over an observational period of 4 years
and using EGM-96, we determined $\mu^{Measured} = 1.1 \pm 0.25 \,$
\cite{ci3}, where 0.25 is the estimated total uncertainty due to
all the error sources --- see Fig. 1.

\begin{figure}[htb]
\includegraphics[width=15cm]{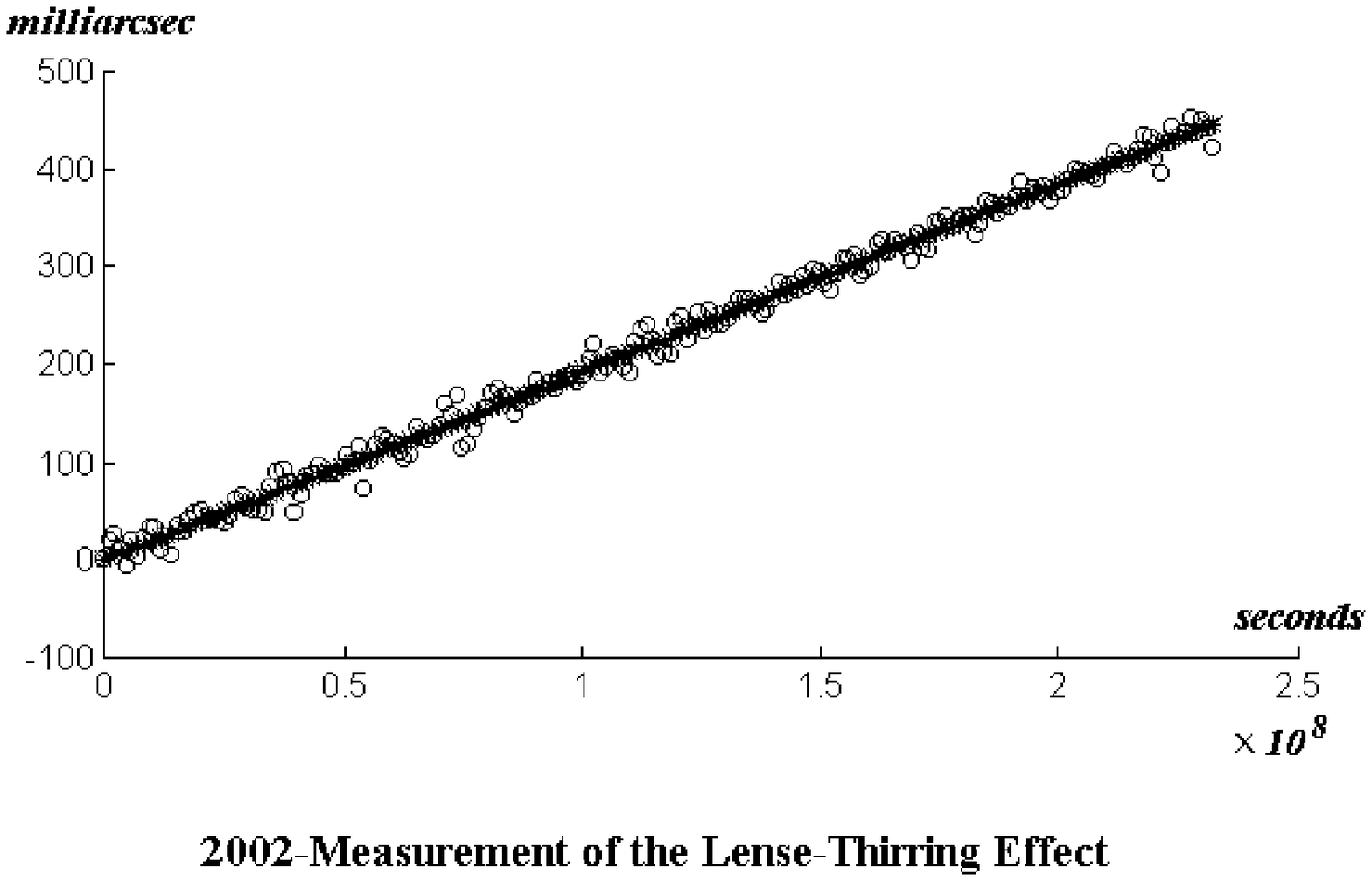}
\caption{\it Latest, 2002, measurement of the Lense-Thirring
effect using LAGEOS and LAGEOS II, obtained by only modeling the
radiation pressure coefficient of LAGEOS II, over nearly 8 years
of data. The best-fit line shown through these combined residuals
has a slope $\mu \simeq$ 1 $\pm$ 0.02. The total estimated
systematic error is about $\pm$ 0.2. The total measured signal is
nearly 440 milliarcsec and the RMS of the post-fit residuals is
about 10 milliarcsec \cite{ci4}.}
\end{figure}

We finally briefly report on our latest, 2002, measurement of the
Lense-Thirring effect over 7.3 years of data of LAGEOS and LAGEOS
II, i.e. over an observational time nearly double than the longer
period of our previous analyses, obtained by only modeling the
radiation pressure coefficient of LAGEOS II (see Fig. 2)
\cite{ci4}.

This recent measurement fully confirms and improves our previous
results: the Lense-Thirring effect exists and its experimental
value, $\mu \cong (1 \pm 0.02) \pm 0.2$ (where $\pm 0.02$ is the
standard deviation of the fit and $\pm 0.2$ is the estimated total
systematic error), fully agrees with the general relativity
prediction of frame-dragging. It is important to notice that: (1)
in the analysis corresponding to Fig. 2 we only modeled on LAGEOS
II the radiation pressure coefficient of the satellite, i.e. the
reflectivity coefficient, $C_R$, and no other parameters such as
the along-track accelerations as in our previous analyses
corresponding to Fig. 1 and ref. \cite{ci2,ci3}; (2) the RMS of
the residuals corresponding to Fig. 2 is about 10 milliarcsec
whereas the total measured signal is about 440 milliarcsec, and
finally (3) the quality of the fit and of the corresponding
measurement can be further improved by further reducing the RMS
of the 15-day fits (corresponding to each point of Fig. 2) with
further processing of the data using GEODYN/SOLVE, thus further
reducing the RMS of the final fit of Fig. 2.

{\it In conclusion, the Lense-Thirring effect exists and its
experimental value, $\mu \cong 1 \pm 0.2$, fully agrees with the
prediction of general relativity \cite{ci4}.}

\end{document}